\documentstyle[12pt]{article}

\advance\textheight by \topskip

\renewcommand{\baselinestretch}{1.2}
\renewcommand{\[}{\begin{eqnarray}}
\renewcommand{\]}{\end{eqnarray}}
\newcommand{\sm}{Standard Model }

\newcounter{hilf}

\begin{document}


\begin{titlepage}
  \renewcommand{\baselinestretch}{1}

  \thispagestyle{empty}
   {\bf \hfill                                       LMU--17/93
   } \\
  \vspace*{1.5cm}
  {\LARGE\bf
  \begin{center}           Top Condensation without Fine--Tuning
  \end{center}  }
  \vspace*{1cm}
  {\begin{center}          {\large Andreas Blumhofer}\footnote{Email:
                           AB at HEP.PHYSIK.UNI--MUENCHEN.DE}
   \end{center} }
  \vspace*{0cm}
  {\sl \begin{center}      Sektion Physik                               \\
                           Ludwig--Maximilians--Universit\"at M\"unchen \\
                           Theresienstr.37, D--80333 M\"unchen
       \end{center} }
  \vspace*{2cm}
  {\Large \bf \begin{center} Abstract  \end{center}  }
  Quadratic divergencies which lead to the usual fine--tuning or
  hierarchy problem are discussed in top condensation models.
  As in the \sm a cancellation of quadratic divergencies is not possible
  without the boson contributions in the radiative corrections which
  are absent in lowest order of an $1/N_c$-expansion. To deal with the
  cancellation of quadratic divergencies we propose therefore an expansion
  in the flavor degrees of freedom. In leading order we find the remarkable
  result that quadratic divergencies automatically disappear.

  \renewcommand{\baselinestretch}{1.2}

\end{titlepage}

  \newpage


\section{Introduction}

Today the \sm is in good agreement with experiment. Even though the Higgs
sector is not yet tested. Due to theoretical problems and from an
aesthetical point of view it is often argued that the \sm Higgs field
cannot be the whole story explaining the symmetry breaking of the
electroweak interaction. The main criticism are the quadratic
divergencies which lead to an unnatural big difference between
bare and renormalized quantities if the Higgs sector is valid up to the
Grand Unification scale. This agrees with 't Hoofts naturalness principle
\cite{Hooft}: A hierarchy problem appears, if there is no extra symmetry
in the case that the Higgs mass goes to zero.

In the \sm quadratic divergencies can be seen in the radiative corrections
of the Higgs tadpoles and self--energies. At the one--loop--level
all massive fermions and bosons contribute with statistically induced
opposite sign
\[     \left( 4m_t^2-M_H^2-M_Z^2-2M_W^2 \right)\Lambda^2,  \label{qd} \]
where light fermion masses are ignored. Therefore the Higgs sector would be
natural, if fermionic and bosonic parts would cancel each other.
This happens for example in supersymmetric theories due to
superpartners\footnote{For a review see \cite{Nilles}.}. In the Standard
Model, however, eq.~(\ref{qd}) does not automatically vanish. To avoid
quadratic divergencies one must postulate eq.~(\ref{qd}) to be zero, which is
the so--called Veltman condition \cite{Veltman}:
\[      4m_t^2 = M_H^2 + M_Z^2 + 2M_W^2.   \label{vm} \]
If $\Lambda$ is very big as suggested above, the vanishing of eq.~(\ref{qd})
needs an extreme fine--tuning, which means that the condition eq.~(\ref{vm})
must be fullfilled very precisely. So far no mechanism or extra symmetry
was found leading in a natural way to the mass relation eq.~(\ref{vm}).
Further more the Veltman condition is regularization scheme dependent, which
can for example be seen by using different cutoffs for different
particles.
Nevertheless the Veltman condition eq.~(\ref{vm}) succeeded in predicting a
top mass bigger than $\sqrt{M_Z^2+2M_W^2}/2 = 73$ GeV and becomes popular
with the increase of the experimental lower bound on the top mass.

For a heavy Higgs and top mass eq.~(\ref{vm}) further degenerates to
\[    M_H \approx 2m_t \; , \label{cc} \]
which suggests that the Higgs boson could be a $\bar{t}t$--boundstate.
It is therefore interesting to ask if top condensation
provides a cancellation mechanism for quadratic divergencies. Keeping
Supersymmetry in mind top and Higgs might play the
role of superpartners in eq.~(\ref{qd}). The term
" Quasi--Supersymmetry " was therefore introduced by Nambu \cite{Nambu}.

But for top condensation models the Higgs boson is a dynamical object
and top and Higgs mass are determined by a complicated system of
Schwinger--Dyson--equations. It seems therefore hard to believe that
both masses are automatically arranged in such a way that
eq.~(\ref{vm}) is fullfilled.
Even more there is no cancellation at all in leading order $1/N_c$
due to the missing next-to-leading order Higgs and Goldstone boson graphs.
To include them one also has to expand in the flavor degrees of freedom,
which we will propose in the second part.
In contrast to the corresponding \sm graphs they get additional factors
leading to the remarkable result that the quadratic divergencies
automatically disappear independent of the values of top and Higgs mass.
In the third part we show further that the cancellation demands a special
group structure of strong and weak interactions which is as an interesting
fact fullfilled by the Standard Model.
In the fourth part we discuss the consequences of the missing quadratic
divergencies for the gap equation and the scale of the new interaction.


\section{Flavor expansion in top condensation}

We start with the minimal top condensation model first discussed
by W.~A.~Bardeen, C.~T.~Hill and M.~Lindner \cite{Bardeen}. It has no
Higgs sector but includes an additional four--fermion--interaction
\[  {\cal L}_I =
    \frac{G}{N_c}\Big(\overline{\psi_L}t_R\Big)\Big(\overline{t_R}\psi_L\Big)
    \;\;;\;\;\;\; \psi_L = {t_L \choose b_L}\;\;. \label{l} \]
But we keep in mind that ${\cal L}_I$ is only an effective low energy
description of some not specified renormalizable
interaction\footnote{For special models see \cite{Hill}.}.
Otherwise higher dimensional operators with arbitrary factors can appear
leading to arbitrary values for top and Higgs mass \cite{Hasenfratz}.

In terms of auxiliary fields $H$, $G^0$ and $G^\pm$ the interaction
has the form
\[   {\cal L}_I &=& \frac{1}{2} \frac{g_t^2N_c}{G}
     \left(HH+G^0G^0+2G^+G^-\right) \nonumber \\ &&
     -\frac{g_t}{\sqrt{2}}H\bar{t}t+i\frac{g_t}{\sqrt{2}}G^0\bar{t}\gamma^5t
     +g_t\left(G^+\bar{t}Lb+G^-\bar{b}Rt\right),
\]
where $L=(1-\gamma^5)/2$ and $R=(1+\gamma^5)/2$.
These fields are arranged in such a way that the usual Higgs and Goldstone
boson couplings to the fermions of the third generation in the \sm appear.

The top mass is now generated by the gap equation (see Fig.~1) which gives
in lowest order $1/N_c$:
\[    1=\frac{2G}{(4\pi)^2}\left(\Lambda^2-m_t^2 \ln\frac{\Lambda^2}{m_t^2}
       \right). \label{ge}   \]
Summing up the geometrical series of the boson vacuum polarisation
of the different auxiliary fields (Fig.~2) and using eq.~(\ref{ge})
one gets propagating boundstates which were introduced above as auxiliary
fields:
\[   D_H &=& \zeta^{-1}\frac{i}{p^2-4m_t^2}  \label{hp} \\
     D_{G^0} &=& \zeta^{-1}\frac{i}{p^2}  \\
     D_{G^\pm} &=& \zeta^{-1}\frac{i}{p^2}
\]
with
\[ \zeta = \frac{N_cg_t^2}{(4\pi)^2}\ln\frac{\Lambda^2}{p^2}. \]
where we have neglected finite terms.
As expected the Higgs mass becomes twice the top mass, which was one of the
motivations for top condensation by comparison with eq.~(\ref{cc}).

The tadpole graph in Fig.~1 is quadratically divergent and corresponds to the
same graph in the Standard Model, which yields the first contribution of
eq.~(\ref{qd}) with one exception:
The Higgs in Fig.~1 is the non propagating auxiliary field, which means that
the Higgs exchange of the tadpole is only a four--fermion--vertex.

One might wonder why there is no Higgs and Goldstone boson tadpole in Fig.~1
with regard to the second contribution in eq.~(\ref{qd}), which is essential
for a possible compensation of quadratic divergencies. But such
graphs would be of higher order in $1/N_c$. Nevertheless motivated by the
\sm one has to expect that the boson tadpoles are of the same order of
magnitude as the top tadpole even if there is no exact cancellation.
This shows that for quadratic divergencies the $1/N_c$--series unfortunately
is a bad expansion in spite of the consistency with regard to
symmetries.

Interestingly the boson contribution increases with the number of Goldstone
bosons and is therefore bigger than in a pure Nambu--Jona--Lasinio model
\cite{Nambu1} containing a U(1) symmetry. That is, the number of Goldstone
bosons depends on the number of particles in the weak multiplet and
therefore on the group structure of the weak interaction. It seems
therefore resonable to choose the group dimension as an additional
expansion parameter. We call this a " flavor expansion " from now on.
For this purpose we generalize the weak $SU(2)_L$ symmetry $  SU(2)_L
\longrightarrow SU(N_L)_L  $. The left--handed doublet $\psi_L$ becomes then:
\[ \psi_L = \left(
   \begin{array}{c} t_L\\b_{1,L}\\ \vdots \\ b_{N_L-1,L}\end{array} \right), \]
where the fields $b_i$ are massless bottom--like fields.
Additional $N_L-1$ massive $W^\pm$--bosons and one massive Z--boson as well as
$N_L(N_L-2)$ massless vector bosons emerge.
All these vector bosons will not play a role in our discussion,
which will become clear in the next section.
As an important fact we have $N_L-1$ charged Goldstone bosons $G^\pm$, which
become the longitudinal components of the $N_L-1$ massive $W^\pm$--bosons
above. In total we end up with the generalized Higgs Lagrangian ${\cal L}_I$:
\[   {\cal L}_I &=& \frac{1}{2} \frac{g_t^2N_c}{G}
     \left(HH+G^0G^0+2\sum_{i=1}^{N_L-1}G_i^+G_i^-\right) \nonumber \\ &&
     -\frac{g_t}{\sqrt{2}}H\bar{t}t+i\frac{g_t}{\sqrt{2}}G^0\bar{t}\gamma^5t
     +g_t\sum_{i=1}^{N_L-1}\left(G_i^+\bar{t}Lb_i+G_i^-\bar{b_i}Rt\right).
\]
Now it seems simple to expand the gap equation and the boson self--energies
in $1/N_c$ and $1/N_L$.
Nevertheless it is not clear what " $1/N_L$ " means. In the following
" Order $(N_L)^n$ " means that there are n summations over the Goldstone boson
degrees of freedom and it does not mean that the factor $N_L$ appears n times.
The reason is that the Higgs boson must also be included with various
combinatorial factors and leads usually to a factor $(N_L+1)^n$ instead of
$(N_L)^n$.

In the next section
we discuss the lowest order in $1/N_c$ and $1/N_L$, which means order
$(1/N_c)^{n_1}\cdot(1/N_L)^{n_2}$ with $n_1+n_2=0$ , where each boson line
contains $1/N_c$ and each flavor summation $N_L$.


\section{The cancellation of quadratic divergencies}

Our aim is now to establish the gap equation in lowest order.
Due to flavor expansion it contains also top self--energies with exchange of
one
or more Higgs or Goldstone bosons like in Fig.~3, where
the two boson lines give $(1/N_c)^2$. Surprisingly there are less than two
flavor summations. Let us consider the first diagram. If the
outer boson line is a $G^+$, the fermion which moves between the outer and
inner boson line is a b--quark. Therefore the inner boson line has to be a
$G^-$ but cannot be a $H$ or $G^0$. That is the reason why there is only one
flavor summation. The second diagram has even no flavor summation.
One comes to the conclusion that only a one Higgs or one Goldstone
boson exchange diagram contributes to the lowest order gap equation.
{}From that it is also not correct to add this one boson diagram to the tadpole
in Fig.~1
because the gap equation would generate the first diagram of Fig.~3 in a
recursive way. One finds that the gap equation must have the form of Fig.~4.
The full top self--energy is denoted by a double circle, which does not
appear in the boson exchange diagram to guarantee the absence of diagrams
of Fig.~3. On the contrary the tadpole diagram is iterated because it contains
the full top self--energy.

It is interesting that the new diagram is not of the order $g_t^2$ as one would
expect with regard to the Standard Model. The boson propagator contains
a factor $\zeta^{-1}$ which is proportional to $g_t^{-2}$ and cancels the
$g_t^2$ coming from the vertices. This is not surprising and expresses the
fact that form factors are of the order one. On the contrary vector bosons
would contribute proportional to $g_1^2$ or $g_2^2$ which really is a higher
order effect.

Unfortunatly the equations of Fig.~4 are a system of coupled integral
equations.
But we are mainly interested in the leading quadratic divergencies of the
tadpole graph. It is therefore possible to combine both equations in the
following way:
We substitute the full double circle in the second equation by the
geometrical series of the first equation denoted with the empty double circle.
But the last graph of the first equation cannot be put in the tadpole twice
or more because the fermion loop of the tadpole graph would be finite due
to more than four outer boson lines. The gap equation for the leading order
divergencies is shown in Fig.~5, which now contains the expected boson
tadpole diagrams as in the \sm yielding the second contribution of
eq.~(\ref{qd}).

In contrast to the usual \sm the boson tadpoles contain an effective vertex
$\Gamma$ with three outer boson lines. As one can easily verify,
in the leading logarithmic divergency there is a simple relation between
$\Gamma$ and the corresponding \sm tree vertex $\Gamma_{SM}$ independent
of the kind of the outer boson lines:
\[    \Gamma = \zeta\left(\frac{2m_t}{M_H}\right)^2 \Gamma_{SM} \]
In the tadpole diagram the $\zeta$ of the three Higgs vertex
is cancelled by the $\zeta^{-1}$ coming from the boson propagator.
Therefore the boson contribution can simply be calculated by using the usual
\sm formula multiplicated by a factor $(2m_t/M_H)^2$.
The two first contributions of eq.~(\ref{qd}) in the \sm give in top
condensation
models:
\[  \left( 4m_t^2- M_H^2\left(\frac{2m_t}{M_H}\right)^2\right)\Lambda^2 \]
which exactly disappear.
The reason for the cancellation is not the relation $M_H=2m_t$ which we have
not demanded but is the fact that the Higgs is a $\bar{t}t$--boundstate and
only couples to the top quark.

The calculation above is not yet a proof to all orders of perturbation theory,
but it seems that
there can be a cancellation mechanism for quadratic divergencies without
Supersymmetry. Even more, this cancellation mechanism depends on the group
structure of the Standard Model.
If we replace
\[  SU(3)_c\times SU(2)_L\times U(1)_Y \to
    SU(N_c)_c\times SU(N_L)_L\times U(1)_Y  \]
the quadratic divergencies read
\[ \left( \frac{N_c}{3} 4 m_t^2-\frac{N_L+1}{3}
   M_H^2\left(\frac{2m_t}{M_H}\right)^2\right)\Lambda^2 \]
and only disappear, if
\[  N_c=N_L+1.  \]
It is interesting that even the \sm has such a structure.
If this mechanism works, it would mean that a \sm with the wrong group
structure would already break at the Grand Unification scale.
But if physics prefers a wide scale hierarchy for example to lower the
energy, the Grand Unified gauge group will prefer to break to the usual
Standard Model.
In this sense top condensation could explain this unusual breaking of the
Grand Unified Theory, which is not understood up to now.

So far we have not included the influence of the vector bosons. As shown
in eq.~(\ref{qd}) they also provide quadratic divergencies in the Standard
Model. In contrast to the Goldstone bosons their couplings to the top
quark are not of the order one but proportional to $g_1$ or $g_2$.
They give higher order contributions to our tadpole diagrams of Fig.~5 and to
the boson self--energy diagrams of Fig.~2, which is shown in Fig.~6.
All corrections can be understood as a redefinition of the
top--Yukawa--coupling.
Therefore the quadratic divergencies of the radiative corrections should
also disappear.
Otherwise gauge bosons would provide a hierarchy problem in contradiction
to 't Hoofts naturalness principle.


\section{The lowest order gap equation}

In the \sm the quadratic divergencies lead to a big
difference between bare and renormalized masses as mentioned above.
In top condensation models the appearance of quadratic divergencies
has a slightly different meaning.
If $\Lambda$ in eq.~(\ref{ge}) is many orders of magnitude larger than the
weak scale, the $m_t/\Lambda$--ratio strongly depends on the choice of the
four--fermion--coupling constant $G$. For a natural range of $G$ $m_t/\Lambda$
is of the order one so that very small values of $m_t/\Lambda$ need an
extreme fine--tuning of $G$.

Including flavor expansion and
after calculating the quadratic and logarithmic divergencies of both graphs
in Fig.~5 the gap equation has the form:
\[ 1=\frac{2G}{(4\pi)^2}\left(\Lambda^2-m_t^2 \ln\frac{\Lambda^2}{m_t^2}
       \right)-\frac{2G}{(4\pi)^2}\left(\Lambda^2-\frac{1}{2}M_H^2
     \ln\frac{\Lambda^2}{m_t^2}\right) \]
or
\[ 1=\frac{2G}{(4\pi)^2}\left(\frac{M_H^2}{2}-m_t^2\right)
     \ln\frac{\Lambda^2}{m_t^2}. \label{ng} \]
The situation changes drastically.
Even for a very large value of $\Lambda$ there is no fine--tuning and the
relation between $G$ and $m_t$ depends only slightly on the cutoff $\Lambda$.
If the Higgs mass is approximately twice the top mass we get
\[ 1=\frac{2G}{(4\pi)^2}m_t^2\ln\frac{\Lambda^2}{m_t^2}. \]
Putting in the Planck mass $\Lambda_{Planck}$ this yields
\[   \frac{2}{(4\pi)^2}\ln\frac{\Lambda_{Planck}^2}{m_t^2}\approx 1  \]
and
\[    G^{-1}\approx m_t^2.    \]
This is a surprising result.
The scale of the interaction which establishes the $\bar{t}t$--boundstate
is of the same order as the top mass itself.
The new interaction could be the \sm itself, or more precisely the
top--Higgs--interaction with a big top--Yukawa--coupling. This is precisely
Nambu's bootstrap idea \cite{Nambu2}, where the Higgs is built up by
interacting with its own constituents.

Nevertheless one has to be careful about equation eq.~(\ref{ng}). We have
only used the leading divergencies which only provide valid results for the
quadratic divergencies.
We must further consider that the flavor expansion is broken by asymmetries
between the Higgs and the Goldstone bosons coming from the different masses.
{}From that one has to expect that the lowest order of the flavor expansion
provides logarithmic divergent radiative corrections to the Goldstone boson
self--energies which unfortunately make them massive. Flavor expansion only
remains a tool to
show the cancellation of quadratic divergencies and cannot be used to
calculate logarithmic corrections to the Higgs and Goldstone boson masses.


\section{Discussion}

Motivated by the similarity between the Veltman condition applied to the usual
\sm and the additional condensate condition for a top mode \sm we have
studied a cancellation mechanism for quadratic divergencies in top
condensation models. In the zeroth order of the usual $1/N_c$--expansion
a cancellation of quadratic divergencies cannot appear due to the missing Higgs
and Goldstone boson contributions.
Using the Goldstone boson degrees of freedom as a further expansion parameter
we could show that the quadratic divergencies cancel to lowest order
without demanding any mass relation and independent of the final values of
the top and the Higgs mass.
This is based on a natural relation between top quark and Higgs boson in
top condensation models similar to the relation between fermions and
bosons in supersymmetry. It is difficult and maybe impossible however to
show such a Quasi--Supersymmetry between a fundamental and a composite
particle in any order of perturbation theory.
Nevertheless it is interesting that the cancellation depends on the group
structure of the weak and strong interaction.
This can be understood in the following way:
We first have to take into account that the Higgs and Goldstone bosons
consists of two constituents. One can therefore replace each boson line
of a Feynman graph by two fermion lines. The interaction between any two
fermions is represented by the small distance of the lines. In this way
the gap equation of fig.~5 can be translated into the form of fig.~7.
The outer top lines have an explicite color index a. One can easily see
that the first graph has neither a color nor a flavor summation. In
contrast the second graph contains a summation over the $N_L+1$ flavor
degrees $t_R^a, t_L^a, b_{1,L}^a,\ldots, b_{N_L-1,L}^a$ ($N_L=2$ in
fig.~7) and an additional factor $(-1)$ due to the extra fermion loop.
The bosons are color neutral so that the inner fermions have the
corresponding color quantum number and no color summation appears.
Even more there is a factor $1/N_c$: After insertion of a
four--fermion--vertex between the outer and inner fermion line there is
no closed color loop but the vertex contains a factor $1/N_c$.
We therefore conclude that the second graph has an additional factor
$-(N_L+1)/N_c$. So in lowest order the quadratic divergencies cancel for
$1-(N_L+1)/N_c=0$. The cancellation is therefore a consequence of
the same number of color and flavor degrees in top
condensation models.
This symmetry between degrees of freedom seems to protect top
condensation models against the presence of quadratic divergencies,
at least in lowest order.

We have also seen that this cancellation cannot be destroyed by the vector
boson contributions due to 't Hoofts naturalness principle.
As one can simply verify following our procedure, the quadratic divergencies
of the boson self--energies also cancel in the same way as the tadpoles.
As above one has to consider the corresponding Standard Model graphs but
must insert effective boson vertices.

We have further shown that the resulting gap equation provides a top mass
of the order of the four--fermion--interaction scale because of the
missing quadratic divergencies.
Since the four--fermion--interaction could be the exchange of Higgs bosons
itself, the bootstrap scenario could be a natural consequence of our
cancellation mechanism.
Nevertheless the experimentell limits on the $\rho$--parameter in the \sm
seems to give a too low top mass and top--Yukawa--coupling, which is not
sufficient to establish a condensation.
In top condensation however the top mass dependence of the $\rho$--parameter
changes and admits larger top masses than in the Standard Model \cite{Lindner}.
Our model could therefore be a real alternative to Supersymmetry but is much
more simple. Its only disadvantage is the difficulty in calculating
dynamical processes beyond the bubble approximation but should not prevent
us from moving on in this direction.

\vspace*{2cm}

Acknowledgements: I would like to thank M.~Beneke, R.~B\"onisch, M.~Lindner,
B.~Stech and L.~Vergara for many useful discussions.


\newpage

{\Large \bf Figure Captions}
\vspace*{0.5cm}

Figure 1: {\sl The gap equation in leading order $1/N_c$. The empty dot is the
one--particle--irreducible self--energy. One gets the full dot, which
represents the full propagator, by summing up the empty dots.}
\\
\\
Figure 2: {\sl Higgs and Goldstone boson self--energies in leading order
$1/N_c$.}
\\
\\
Figure 3: {\sl Higher order contributions to the top self--energy in
color and flavor expansion.}
\\
\\
Figure 4: {\sl The system of gap equations in leading order $1/N_c$
and $1/N_L$. The full double circle is the full top propagator, which
can be found by summing up the empty double circle. The full and empty
dots are auxiliary objects and the full dot is the sum of the
empty one.}
\\
\\
Figure 5: {\sl The gap equation in leading order $1/N_c$ and $1/N_L$ and
in the leading divergencies.}
\\
\\
Figure 6: {\sl Vector boson contributions to the gap equation.}
\\
\\
Figure 7: {\sl A form of the gap equation of Fig.~5, where the constituents
are visible. It shows the group structure dependence of the cancellation
mechanism.}


\end{document}